\documentclass[11pt]{article}
\makeatletter\renewcommand{\section}{\@startsection
{section}{1}{\z@}{-3.5ex plus -1ex minus
    -.2ex}{2.3ex plus .2ex}{\bf }}

\makeatletter\renewcommand{\subsection}{\@startsection{subsection}{2}{\z@}{-3.25ex
plus -1ex minus
   -.2ex}{1.5ex plus .2ex}{\it }}
\makeatletter\renewcommand{\subsubsection}{\@startsection{subsubsection}{3}{-2.45ex}{-3.25ex
plus -1ex minus -.2ex}{1.5ex plus .2ex}{\it }}

\renewcommand{\thesubsection}{\arabic{section}.\arabic{subsection}.}

\renewcommand{\theequation}{\thesection\arabic{equation}}

\let\fn\footnote
\renewcommand{\footnote}[1]{\linespread{1.1}\fn{#1}\linespread{1.29}}

\usepackage[english]{babel}
\usepackage{amsmath,amssymb}
\usepackage{bbm}

\def\slasha#1{\setbox0=\hbox{$#1$}#1\hskip-\wd0\hbox to\wd0{\hss\sl/\/\hss}}

\def\periodb#1{\setbox0=\hbox{$#1$}#1\hskip-\wd0\hbox to\wd0{-}}
\newcommand{\lsc}{\{\hspace{-0.1cm}[}
\newcommand{\rsc}{]\hspace{-0.1cm}\}}
\newcommand{\unit}{\mathbbm{1}}   
\newcommand{\id}{\mathrm{id}}   
\newcommand{\CF}{\mathcal{F}}    
\newcommand{\CH}{\mathcal{H}}    
\newcommand{\CL}{\mathcal{L}}    
\newcommand{\CN}{\mathcal{N}}    
\newcommand{\CO}{\mathcal{O}}    
\newcommand{\CS}{\mathcal{S}}    
\newcommand{\CU}{\mathcal{U}}    
\newcommand{\CW}{\mathcal{W}}    
\newcommand{\frg}{\mathfrak{g}}    
\newcommand{\fra}{\mathfrak{a}}    
\newcommand{\FR}{\mathbbm{R}}     
\newcommand{\FC}{\mathbbm{C}}     
\newcommand{\RZ}{\mathbbm{Z}}     
\newcommand{\dd}{\mathrm{d}}     
\newcommand{\dpar}{\partial}     
\newcommand{\de}{\mathrm{e}}     
\newcommand{\di}{\mathrm{i}}     
\newcommand{\btheta}{{\bar{\theta}}}     
\newcommand{\bl}{{\bar{\lambda}}}     
\newcommand{\ald}{{\dot{\alpha}}}     
\newcommand{\bed}{{\dot{\beta}}}     
\newcommand{\gad}{{\dot{\gamma}}}     
\newcommand{\eps}{{\varepsilon}}     
\newcommand{\eand}{{~~~\mbox{and}~~~}}     
\newcommand{\der}[1]{\frac{\dpar}{\dpar #1}}   
\newcommand{\dder}[1]{\frac{\dd}{\dd #1}}   

\newcommand{\sU}{\mathsf{U}}     
\newcommand{\ssp}{\FR^{4|4\CN}}     
\newcommand{\sspdef}{\FR^{4|4\CN}_\hbar}     
\newcommand{\remark}[1]{}     

\makeatletter \@addtoreset{equation}{section} \makeatother

\renewcommand{\theequation}{\thesection\arabic{equation}}

\begin{document}
\begin{titlepage}
\setcounter{page}{0}
\begin{flushright}
{\tt hep-th/0506057}\\
UTTG--04--05\\
ITP--UH--09/05\\
\end{flushright}
\vskip 0.7cm
\begin{center}
{\LARGE \bf Drinfeld-Twisted Supersymmetry and\\[0.5cm] Non-Anticommutative
Superspace} \vskip 1.0cm {\large Matthias Ihl$^1$ and Christian
S\"{a}mann$^2$} \vskip 1cm
{\em ${}^1$Department of Physics\\
University of Texas at Austin\\
Austin, TX 78712, USA\\
{\tt msihl@zippy.ph.utexas.edu}}\\[5mm]
{\em ${}^2$ Institut f\"{u}r Theoretische Physik\\
Universit\"{a}t Hannover\\
Appelstra{\ss}e 2, D-30167 Hannover, Germany\\
{\tt saemann@itp.uni-hannover.de}}\\[5mm]
\vskip 0.5cm
\end{center}
\begin{center}
{\bf Abstract}
\end{center}
\begin{quote}
We extend the analysis of {\tt hep-th/0408069} on a Lorentz
invariant interpretation of noncommutative spacetime to field
theories on non-anticommutative superspace with half the
supersymmetries broken. By defining a Drinfeld-twisted Hopf
superalgebra, it is shown that one can restore {\em twisted}
supersymmetry and therefore obtain a twisted version of the chiral
rings along with certain Ward-Takahashi identities. Moreover, we
argue that the representation content of theories on the deformed
superspace is identical to that of their undeformed cousins and
comment on the consequences of our analysis concerning
non-renormalization theorems. \vskip 25mm {\footnotesize
}
\end{quote}
\end{titlepage}

\section{Introduction}

Over the last decade, there has been an immense effort by string
theorists to improve our understanding of string dynamics in
nontrivial backgrounds. Most prominently, Seiberg and Witten
\cite{Seiberg:1999vs} discovered that superstring theory in a
constant Kalb-Ramond 2-form background can be formulated in terms
of field theories on noncommutative spacetimes upon taking the
so-called Seiberg-Witten zero slope limit. Subsequently, these
noncommutative variants of ordinary field theories were intensely
studied, revealing many interesting new aspects, such as UV--IR
mixing \cite{Minwalla:1999px}, the vastness of nontrivial
classical solutions to the field equations\footnote{Many classical
solutions were obtained with the help of solution generating
techniques, generalized to the noncommutative setting in
\cite{Nekrasov:2000zz}-\cite{Domrin:2004pg}.}
\cite{Konechny:2000dp,Douglas:2001ba,Gross:2000ss} and the
nonsingular nature of instanton moduli spaces, see e.g.\
\cite{Nekrasov:2000zz}. It turned out that as low energy effective
field theories, noncommutative field theories exhibit many
manifestations of stringy features descending from the underlying
string theory. Therefore, these theories have proven to be an
ideal toy model for studying string theoretic questions which
otherwise remain intractable, e.g.\ tachyon condensation
\cite{Sen:1998rg}-\cite{Kajiura:2001pq} and further dynamical
aspects of strings \cite{Gross:2000ph} (for recent work, see e.g.\
\cite{Wimmer:2005bz, Popov:2005ik}).

Expanding on the analysis of \cite{Ooguri:2003qp}, Seiberg
\cite{Seiberg:2003yz} showed that there is a deformation of
Euclidean $\CN=1$ superspace in four dimensions which leads to a
consistent supersymmetric field theory with half of the
supersymmetries broken. The idea was to deform the algebra of the
anticommuting coordinates $\theta$ to the Clifford algebra
\begin{equation}
\{\theta^{\alpha},\theta^{\beta}\}=C^{\alpha,\beta}~,
\end{equation}
which arises from considering string theory in a background with a
constant graviphoton field strength. For earlier work in this
area, see \cite{Schwarz:1982pf}-\cite{deBoer:2003dn}. Later on,
non-anticommutativity for extended supersymmetry was considered,
as well \cite{Ivanov:2003te,Ferrara:2003xk,Saemann:2004cf}. These
discoveries triggered many publications studying
non-anticommutative field theories. Of particular interest to our
work is the question of renormalizability of non-anticommutative
field theories, e.g.\ of the $\CN=\frac{1}{2}$ Wess-Zumino model,
as discussed in \cite{Terashima:2003ri}-\cite{Alishahiha:2003kg}.

Another recent development began with the realization that
noncommutative field theories, although manifestly breaking
Poincar{\'e} symmetry,\footnote{In noncommutative spacetime, the
Poincar{\'e} group is broken down to the stabilizer subgroup of the
deformation tensor.} can be recast into a form which is invariant
under a twist-deformed action of the Poincar{\'e} algebra
\cite{Oeckl:2000eg,Chaichian:2004za,Chaichian:2004yh}. In this
framework, the commutation relation $[x^{\mu},x^{\nu}]=\di
\Theta^{\mu \nu}$ is understood as a result of the non-{\it
co}commutativity of the coproduct of a twisted Hopf Poincar{\'e}
algebra acting on the coordinates. This result can be used to show
that the representation content of Moyal-Weyl-deformed theories is
identical to that of their undeformed Lorentz invariant
counterparts. Furthermore, theorems in quantum field theory which
require Lorentz invariance for their proof can now be carried over
to the Moyal-Weyl-deformed case using twisted Lorentz invariance.
For related works, see
\cite{Banerjee:2004ev}-\cite{Aschieri:2005yw}.

The purpose of the present paper is to extend the analysis of
\cite{Chaichian:2004za,Chaichian:2004yh} to supersymmetric field
theories on non-anticommutative superspaces and use
Dinfeld-twisted supersymmetry to translate properties of these
field theories into the non-anticommutative situation, where half
of the supersymmetries are broken. Note that twisted supersymmetry
was already considered in \cite{Kobayashi:2004ep}. However, the
analysis of extended supersymmetries presented in this reference
differs from the one we will propose here.

The paper is organized as follows: We will fix our conventions for
non-anticommutative superspace in section 2 and introduce the
Drinfeld twist of the Euclidean super Poincar{\'e} algebra and its
universal enveloping algebra in section 3. Then, in section 4, we
will discuss the implications of these mathematical structures
concerning the representation content, the reemergence of
(twisted) chiral rings and Ward-Takahashi identities. Moreover, we
comment on non-renormalization theorems in the twisted
supersymmetric case before we conclude in section 5. Some basic
definitions and a useful extension of the Baker-Campbell-Hausdorff
formula can be found in the appendix.

\section{Non-anticommutative superspace}\label{sec:nacss}

\subsection{Superspace conventions}

Throughout this paper, we will mostly adopt the conventions of
\cite{Seiberg:2003yz}. Consider the four-dimensional Euclidean
space $\FR^4$ with coordinates $(x^\mu)$ and extend it to the
space $\FR^{4|4\CN}$ by adding $4\CN$ Gra{\ss}mann coordinates
$(\theta^{i\alpha},\btheta^\ald_i)$ with\footnote{Strictly
speaking, this superspace with Euclidean signature can be
consistently defined only for an even number of supersymmetries,
as the appropriate reality condition for $\theta^{i\alpha}$ and
$\btheta^\ald_i$ is a symplectic Majorana condition and
establishes a pairwise relation between these spinors. When
working on the complexified superspace $\FC^{4|4\CN}$, i.e., when
``temporarily doubling the fermionic degrees of freedom'', this
obstacle however disappears.} $i=1,\ldots,\CN$ and
$\alpha,\ald=1,2$. The algebra of superfunctions on this space is
denoted by
\begin{equation}
\CS\ :=\ C^\infty\otimes\Lambda_{4\CN}~,
\end{equation}
where $\Lambda_{4\CN}:=\Lambda^\bullet(\FR^{4\CN})$ is the
Gra{\ss}mann algebra with $4\CN$ generators. As it is well known, an
element of $\CS$ can be decomposed into its Gra{\ss}mann even and its
Gra{\ss}mann odd part as well as into its ``body'' (the purely bosonic
part) and its ``soul'' (the nilpotent part), cf.\ e.g.\
\cite{Cartier:2002zp}.

Recall that translations in the Gra{\ss}mann directions of this space
are generated by the superderivatives and the supercharges which
act on a superfunction $f\in \CS$ as
\begin{equation}
\begin{aligned}
D_{i\alpha}f&\ :=\
\der{\theta^{i\alpha}}f+\di\sigma^\mu_{\alpha\ald}\btheta^\ald_i\dpar_\mu
f~,&\bar{D}^i_{\ald}f&\ :=\
-\der{\btheta_i^\ald}f-\di\theta^{i\alpha}\sigma^\mu_{\alpha\ald}\dpar_\mu
f~,\\
Q_{i\alpha}f&\ :=\
\der{\theta^{i\alpha}}f-\di\sigma^\mu_{\alpha\ald}\btheta^\ald_i\dpar_\mu
f~,&\bar{Q}^i_{\ald}f&\ :=\
-\der{\btheta_i^\ald}f+\di\theta^{i\alpha}\sigma^\mu_{\alpha\ald}\dpar_\mu
f~,
\end{aligned}
\end{equation}
and satisfy the algebra (we do not allow for central charges)
\begin{equation}
\begin{aligned}
\{D_{i\alpha},D_{j\beta}\}&\ =\ 0~,&
\{\bar{D}^i_{\ald},\bar{D}^j_{\bed}\}&\ =\ 0~,&
\{D_{i\alpha},\bar{D}^j_{\bed}\}&\ =\ -2\di\delta_i^j\sigma^\mu_{\alpha\bed}\dpar_\mu~,\\
\{Q_{i\alpha},Q_{j\beta}\}&\ =\ 0~,&
\{\bar{Q}^i_{\ald},\bar{Q}^j_{\bed}\}&\ =\ 0~,&
\{Q_{i\alpha},\bar{Q}^j_{\bed}\}&\ =\
2\di\delta_i^j\sigma^\mu_{\alpha\bed}\dpar_\mu~.
\end{aligned}
\end{equation}
Our discussion simplifies considerably if we switch to the
left-handed chiral coordinates
\begin{equation}
(y^\mu:=x^\mu+\di
\theta^{i\alpha}\sigma^\mu_{\alpha\ald}\btheta^\ald_i,~
\theta^{i\alpha},~\btheta^\ald_i)~,
\end{equation}
in which the representations of the superderivatives and the
supercharges read
\begin{equation}
\begin{aligned}
D_{i\alpha}f&\ =\
\der{\theta^{i\alpha}}f+2\di\sigma^\mu_{\alpha\ald}\btheta^\ald_i\dpar^L_\mu
f~,&\bar{D}^i_{\ald}f&\ =\ -\der{\btheta^\ald_i}f~,\\
Q_{i\alpha}f&\ =\ \der{\theta^{i\alpha}}f~,&\bar{Q}^i_{\ald}f&\ =\
-\der{\btheta^\ald_i}f+2\di\theta^{i\alpha}\sigma^\mu_{\alpha\ald}\dpar^L_\mu
f~,
\end{aligned}
\end{equation}
where $\dpar^L_\mu$ denotes a derivative with respect to
$y^\mu$. Due to $\dpar^L_\mu=\dpar_\mu$, we drop the superscript
``$L$'' in the following.

\subsection{The canonical non-anticommutative deformation}

The canonical deformation of $\ssp$ to $\sspdef$ amounts to
\cite{Seiberg:2003yz}
\begin{equation}
\{\hat{\theta}^{i\alpha},\hat{\theta}^{j\beta}\}\ =\ \hbar
C^{i\alpha,j\beta}~,
\end{equation}
where the hats indicate, as usual, that we are dealing with an
operator representation. Equivalently, one can instead deform the
algebra of superfunctions $\CS$ on $\ssp$ to an algebra
$\CS_\star$, in which the product is given by the Moyal-type star
product
\begin{equation}\label{candef}
f\star g\ =\ f\exp\left(-\frac{\hbar}{2}
\overleftarrow{Q}_{i\alpha}
C^{i\alpha,j\beta}\overrightarrow{Q}_{j\beta}\right) g~,
\end{equation}
where $\overleftarrow{Q}_{i\alpha}$ and
$\overrightarrow{Q}_{j\beta}$ are supercharges acting from the
right and the left, respectively. Recall that
$\theta^{i\alpha}\overleftarrow{Q}_{j\beta}=-\delta^i_j\delta_\beta^\alpha$.
All commutators involving this star multiplication will be denoted
by a $\star$, e.g.\ the graded commutator will read
\begin{equation}
\lsc f,g\rsc_\star\ :=\ f\star g-(-1)^{\tilde{f}\tilde{g}}g\star
f~.
\end{equation}
with $\tilde{f}$ and $\tilde{g}$ denoting the grading of $f$ and
$g$, respectively, cf.\ appendix A.

In chiral coordinates, we have the following coordinate algebra on
$\CS_\star$:
\begin{equation}\label{candefalgebra}
\begin{aligned}
&\hspace{3.5cm}\{\theta^{i\alpha},\theta^{j\beta}\}_\star\ =\
\hbar C^{i\alpha,j\beta}~,\\ &[y^\mu,y^\nu]_\star\ =\
[y^\mu,\theta^{i\alpha}]_\star\ =\ [y^\mu,\theta^{i\alpha}]_\star\
=\ \{\theta^{i\alpha},\btheta^\ald_i\}_\star\ =\
\{\btheta^\ald_i,\btheta^\bed_j\}_\star\ =\ 0~.
\end{aligned}
\end{equation}
This deformation has been shown to arise in string theory from
open superstrings of type IIB in the background of a constant
graviphoton field strength
\cite{Ooguri:2003qp,Seiberg:2003yz,deBoer:2003dn}. The
corresponding deformed algebra of superderivatives and
supercharges reads
\begin{equation}\label{defalgebra}
\begin{aligned}
\{D_{i\alpha},D_{j\beta}\}_\star&\ =\ 0~,~~~
\{\bar{D}^i_{\ald},\bar{D}^j_{\bed}\}_\star\ =\ 0~,\\
\{D_{i\alpha},\bar{D}^j_{\bed}\}_\star&\ =\
-2\di\delta_i^j\sigma^\mu_{\alpha\bed}\dpar_\mu~,\\
\{Q_{i\alpha},Q_{j\beta}\}_\star&\ =\ 0~,~~~
\{\bar{Q}^i_{\ald},\bar{Q}^j_{\bed}\}_\star\ =\ -4\hbar
C^{i\alpha,j\beta}\sigma_{\alpha\ald}^\mu\sigma_{\beta\bed}^\nu\dpar_\mu\dpar_\nu~,\\
\{Q_{i\alpha},\bar{Q}^j_{\bed}\}_\star&\ =\
2\di\delta_i^j\sigma^\mu_{\alpha\bed}\dpar_\mu~.
\end{aligned}
\end{equation}
By inspection of this deformed algebra, it becomes clear that the
number of supersymmetries is reduced to $\CN/2$, since those
generated by $\bar{Q}^i_\ald$ will be broken.\footnote{Note that
this result is due to the fact that we are using Euclidean
spacetime. In Minkowski superspace, $Q$ and $\bar{Q}$ are related
via complex conjugation, and it is therefore not possible to break
only half of the supersymmetries.} On the other hand, it still
allows for the definition of chiral and anti-chiral superfields as
the algebra of the superderivatives $D_{i\alpha}$ and
$\bar{D}^i_\ald$ is undeformed.

An alternative approach, which was followed in
\cite{Ferrara:2000mm}, manifestly preserves supersymmetry but
breaks chirality. This simply means that one replaces the
supercharges $Q_{i\alpha}$ by the superderivatives $D_{i\alpha}$
in the definition of the deformation \eqref{candef}. Without
chiral superfields, however, it is impossible to define super
Yang-Mills theory in the standard superspace formalism.

In the approach we will present in the following, supersymmetry
{\em and} chirality are manifestly and simultaneously preserved,
albeit in a twisted form.

\section{Drinfeld twist of the Euclidean super Poincar{\'e} algebra}\label{sec:drinfeld}

\subsection{The Euclidean super Poincar{\'e} algebra and its enveloping algebra}

The starting point of our discussion is the ordinary Euclidean
super Poincar{\'e} algebra\footnote{or inhomogeneous super Euclidean
algebra} $\frg$ on $\ssp$ without central extensions, which
generates the isometries on the space $\ssp$. More explicitly, we
have the generators of translations $P_\mu$, the generators of
four-dimensional rotations $M_{\mu\nu}$ and the $4\CN$
supersymmetry generators $Q_{i\alpha}$ and $\bar{Q}_\ald^i$. They
satisfy the following algebra:
\begin{equation}\label{sepalgebra}
\begin{aligned}
\begin{aligned}
{}[P_\rho,M_{\mu\nu}]&\ =\
\di(\delta_{\mu\rho}P_\nu-\delta_{\nu\rho}P_\mu)~,\\
{}[M_{\mu\nu},M_{\rho\sigma}]&\ =\
-\di(\delta_{\mu\rho}M_{\nu\sigma}-\delta_{\mu\sigma}M_{\nu\rho}-
\delta_{\nu\rho}M_{\mu\sigma}+\delta_{\nu\sigma}M_{\mu\rho})~,
\end{aligned}\hspace{0.65cm}\\
\begin{aligned}
{}[P_\mu,Q_{i\alpha}]&\ =\ 0~,&[P_\mu,\bar{Q}_\ald^i]&\ =\ 0~,&\\
[M_{\mu\nu},Q_{i \alpha}]&\ =\ \di
(\sigma_{\mu\nu})_\alpha{}^\beta
Q_{i\beta}~,&[M_{\mu\nu},\bar{Q}^{i\ald}]&\ =\ \di
(\bar{\sigma}_{\mu\nu})^\ald{}_\bed
\bar{Q}^{i\bed}~,\\
\{Q_{i\alpha},\bar{Q}^j_\bed\}&\ =\
2\delta_i^j\sigma_{\alpha\bed}^\mu P_\mu
~,&\{Q_{i\alpha},Q_{j\beta}\}&\ =\
\{\bar{Q}^i_{\ald},\bar{Q}^j_\bed\}=0~.
\end{aligned}
\end{aligned}
\end{equation}
The Casimir operators of the Poincar{\'e} algebra used for labelling
representations are $P^2$ and $W^2$, where the latter is the
square of the Pauli-Ljubanski operator
\begin{equation}
W_{\mu}\ =\ -\tfrac{1}{2} \epsilon_{\mu \nu \rho\sigma} M^{\nu
\rho} P^{\sigma}~.
\end{equation}
This operator is, however, not a Casimir of the super Poincar{\'e}
algebra; instead, there is a supersymmetric variant: the (superspin)
operator $\widetilde{C}^2$ defined as the square of
\begin{equation}
\widetilde{C}_{\mu \nu}\ =\ \widetilde{W}_{\mu} P_{\nu} -
\widetilde{W}_{\nu} P_{\mu},
\end{equation}
where $\widetilde{W}_{\mu}:= W_{\mu}- \frac{1}{4} \bar{Q}^i_{\ald}
\bar{\sigma}^{\ald \alpha}_{\mu}Q_{i \alpha}$.

Recall that a {\em universal enveloping algebra} $\CU(\fra)$ of a
Lie algebra $\fra$ is an associative unital algebra together with
a Lie algebra homomorphism $h:\fra\rightarrow \CU(\fra)$,
satisfying the following universality property: For any further
associative algebra $A$ with homomorphism $\phi:\fra\rightarrow
A$, there exists a unique homomorphism $\psi:\CU(\fra)\rightarrow
A$ of associative algebras, such that $\phi=\psi\circ h$. Every
Lie algebra has an universal enveloping algebra, which is unique
up to algebra isomorphisms.

The univeral enveloping algebra $\CU(\frg)$ of the Euclidean super
Poincar{\'e} algebra $\frg$ is a cosupercommutative Hopf
superalgebra\footnote{cf.\ appendix} with counit and coproduct
defined by $\eps(\unit)=1$ and $\eps(x)=0$ otherwise,
$\Delta(\unit)=\unit\otimes \unit$ and $\Delta(x)=\unit\otimes
x+x\otimes \unit$ otherwise.

\subsection{The Drinfeld twist of the enveloping algebra}

Given a Hopf algebra $H$ with coproduct $\Delta$, a counital
2-cocycle $\CF$ is a counital element of $H\otimes H$, which has
an inverse and satisfies
\begin{equation}\label{cocyclecond}
\CF_{12}(\Delta\otimes\id)\CF\ =\ \CF_{23}(\id\otimes\Delta)\CF~,
\end{equation}
where we used the common shorthand notation
$\CF_{12}=\CF\otimes\unit$, $\CF_{23}=\unit\otimes\CF$ etc. As
done in \cite{Chaichian:2004za}, such a counital 2-cocyle $\CF\in
H\otimes H$ can be used to define a twisted Hopf
algebra\footnote{This twisting amounts to constructing a
quasitriangular Hopf algebra, as discussed, e.g., in
\cite{Chari:1994pz}.} $H^\CF$ with a new coproduct given by
\begin{equation}
\Delta^\CF(Y)\ :=\ \CF\Delta(Y)\CF^{-1}~.
\end{equation}
The element $\CF$ is called a {\em Drinfeld twist}; such a
construction was first considered in \cite{Drinfeld:1989st}.

For our purposes, i.e.\ to recover the canonical algebra of
non-anticommutative coordinates \eqref{candefalgebra}, we choose
the abelian twist $\CF\in\CU(\frg)\otimes\CU(\frg)$ defined by
\begin{equation}
\CF\ =\ \exp\left(-\frac{\hbar}{2} C^{i\alpha,j\beta}
Q_{i\alpha}\otimes Q_{j\beta}\right)~.
\end{equation}
As one easily checks, $\CF$ is indeed a counital 2-cocycle: First,
it is invertible and its inverse is given by
$\CF^{-1}=\exp\left(\frac{\hbar}{2} C^{i\alpha,j\beta}
Q_{i\alpha}\otimes Q_{j\beta}\right)$. (Because the $Q_{i\alpha}$
are nilpotent, $\CF$ and $\CF^{-1}$ are not formal series but
rather finite sums.) Second, $\CF$ is counital since it satisfies
the conditions
\begin{equation}
(\eps\otimes \id)\CF\ =\ \unit\eand(\id\otimes \eps)\CF\ =\
\unit~,
\end{equation}
as can be verified without difficulty. Also, the remaining cocycle condition
\eqref{cocyclecond} turns out to be fulfilled since
\begin{equation}
\begin{aligned}
\CF_{12}(\Delta\otimes\id)\CF&\ =\
\CF_{12}\exp\left(-\frac{\hbar}{2}
C^{i\alpha,j\beta}(Q_{i\alpha}\otimes \unit+\unit\otimes
Q_{i\alpha})\otimes Q_{j\beta}\right)~,\\
\CF_{23}(\id\otimes\Delta)\CF&\ =\
\CF_{23}\exp\left(-\frac{\hbar}{2}
C^{i\alpha,j\beta}Q_{i\alpha}\otimes(Q_{j\beta}\otimes
\unit+\unit\otimes Q_{j\beta})\right)
\end{aligned}
\end{equation}
yields, due to the commutativity of the $Q_{i\alpha}$,
\begin{equation}
\CF_{12}\CF_{13}\CF_{23}\ =\ \CF_{23}\CF_{12}\CF_{13}~,
\end{equation}
which is obviously true.

Note that after introducing this Drinfeld twist, the
multiplication in $\CU(\frg)$ and the action of $\frg$ on the
coordinates remain the same. In particular, the representations of
the twisted and the untwisted algebras are identical. It is only
the action of $\CU(\frg)$ on the tensor product of the
representation space, given by the coproduct, which changes.

Let us be more explicit on this point: the coproduct of the
generator $P_\mu$ does not get deformed, as $P_\mu$ commutes with
$Q_{j\beta}$:
\begin{equation}\label{eq:P}
\Delta^\CF(P_\mu)\ =\ \Delta(P_\mu)~.
\end{equation}
For the other generators of the Euclidean super Poincar{\'e} algebra,
the situation is slightly more complicated. Due to the
rule\footnote{cf.\ appendix} $(a_1\otimes a_2)(b_1\otimes
b_2)=(-1)^{\tilde{a}_2\tilde{b}_1}(a_1b_1\otimes a_2 b_2)$, where
$\tilde{a}$ denotes the Gra{\ss}mann parity of $a$, we have the
relations\footnote{Here, $I_k$ and $J_k$ are multi-indices, e.g.\
$I_k=i_k \alpha_k$.} (cf.\ equation \eqref{bchformula})
\begin{align}\nonumber
&\CF \left(D\otimes \unit\right)\CF^{-1}\ =\
\\\nonumber&\hspace{1cm} \sum_{n=0}^\infty
\frac{(-1)^{n\tilde{D}+\frac{n(n-1)}{2}}}{n!}\left(-\frac{\hbar}{2}\right)^nC^{I_1J_1}\ldots
C^{I_nJ_n}\lsc Q_{I_1},\lsc\ldots\lsc Q_{I_n},D\rsc\rsc\rsc
\otimes Q_{J_1}\ldots Q_{J_n}~,\\ &\CF \left(\unit\otimes
D\right)\CF^{-1}\ =\ \\&\hspace{1cm} \sum_{n=0}^\infty
\frac{(-1)^{n\tilde{D}+\frac{n(n-1)}{2}}}{n!}\left(-\frac{\hbar}{2}\right)^nC^{I_1J_1}\ldots
C^{I_nJ_n} Q_{I_1}\ldots Q_{I_n}\otimes \lsc
Q_{J_1},\lsc\ldots\lsc Q_{J_n},D\rsc\rsc\rsc~,\nonumber
\end{align}
where $\lsc\cdot,\cdot\rsc$ denotes the graded commutator. {}From
this, we immediately obtain
\begin{equation}
\Delta^\CF(Q_{i\alpha})\ =\ \Delta(Q_{i\alpha})~.
\end{equation}
Furthermore, we can also derive the expressions for
$\Delta^\CF(M_{\mu\nu})$ and $\Delta^\CF(\bar{Q}_\gad^k)$, which
read
\begin{align}
\Delta^\CF(M_{\mu\nu}) &\ =\ \Delta(M_{\mu\nu})+\frac{\di
\hbar}{2} C^{i\alpha,j\beta} \left[
(\sigma_{\mu\nu})_\alpha{}^\gamma Q_{i\gamma} \otimes
Q_{j\beta}+Q_{i\alpha} \otimes (\sigma_{\mu\nu})_\beta{}^\gamma
 Q_{j\gamma}\right]~, \\
\Delta^\CF(\bar{Q}^k_{\gad})&\ =\ \Delta(\bar{Q}^k_{\gad}) + \hbar
C^{i\alpha,j\beta} \left[ \delta^k_i\sigma_{\alpha\gad}^\mu P_\mu
\otimes Q_{j\beta}+Q_{i\alpha} \otimes \delta^k_j
\sigma_{\beta\gad}^\mu P_\mu \right].
\end{align}
The twisted coproduct of the Pauli-Ljubanski operator $W_{\mu}$
becomes
\begin{equation}\label{eq:Wtwist}
\Delta^\CF(W_{\mu})\ =\ \Delta(W_{\mu}) -\frac{\di \hbar}{4}
C^{i\alpha , j \beta} \epsilon_{\mu \nu \rho\sigma} \left( Q_{i
\alpha} \otimes (\sigma^{\nu \rho})_{\beta}{}^{\gamma} Q_{j
\gamma} P^{\sigma} + (\sigma^{\nu \rho})_{\alpha}{}^{\gamma} Q_{i
\gamma} P^{\sigma} \otimes Q_{j \beta} \right)~,
\end{equation}
while for its supersymmetric variant $\widetilde{C}_{\mu\nu}$, we
have
\begin{equation}\label{eq:C2}
\begin{aligned}
\Delta^\CF(\widetilde{C}_{\mu\nu}) &\ =\
\Delta(\widetilde{C}_{\mu\nu}) -\frac{\hbar}{2} C^{i\alpha , j
\beta} \left[ Q_{i\alpha} \otimes Q_{j \beta},
\Delta(\widetilde{C}_{\mu\nu})\right] \\ &\ =\
\Delta(\widetilde{C}_{\mu\nu})- \frac{\hbar}{2} C^{i\alpha , j
\beta} \left( \left[Q_{i\alpha},\widetilde{C}_{\mu\nu}\right]
\otimes Q_{j \beta} + Q_{i \alpha} \otimes
\left[Q_{j \beta},\widetilde{C}_{\mu\nu}\right]\right)\\
&\ =\ \Delta(\widetilde{C}_{\mu\nu})~,
\end{aligned}
\end{equation}
since $[Q_{i \alpha},\widetilde{C}_{\mu\nu}]=0$ by construction.

\subsection{Representation on the algebra of superfunctions}

Given a representation of the Hopf algebra $\CU(\frg)$ in an
associative algebra consistent with the coproduct $\Delta$, one
needs to adjust the multiplication law after introducing a
Drinfeld twist. If $\CF^{-1}$ is the inverse of the element
$\CF\in\CU(\frg)\otimes \CU(\frg)$ generating the twist, the new
product compatible with $\Delta^\CF$ reads
\begin{equation}\label{multiplication}
a\star b\ :=\ m^\CF (a\otimes b)\ :=\ m\circ \CF^{-1}(a\otimes
b)~,
\end{equation}
where $m$ denotes the ordinary product $m(a\otimes b)=ab$.

Let us now turn to the representation of the Hopf superalgebra
$\CU(\frg)$ on the algebra
$\CS:=C^{\infty}(\FR^4)\otimes\Lambda_{4\CN}$ of superfunctions on
$\FR^{4|4\CN}$. On $\CS$, we have the standard representation of
the super Poincar{\'e} algebra in chiral coordinates
$(y^\mu,\theta^{i\alpha},\btheta^\ald_i)$:
\begin{equation}
\begin{aligned}
P_\mu f&\ =\ \di\dpar_\mu f~,
&M_{\mu\nu}f&\ =\ \di(y_\mu\dpar_\nu-y_\nu\dpar_\mu)f~,\\
Q_{i\alpha}f&\ =\ \der{\theta^{i\alpha}}f~,& \bar{Q}^i_\ald f&\ =\
\left(-\der{\btheta^\ald_i}f+2\di\theta^{i\alpha}\sigma^\mu_{\alpha\ald}\dpar_\mu\right)
f~,
\end{aligned}
\end{equation}
where $f$ is an element of $\CS$. After the twist, the
multiplication $m$ becomes the twist-adapted multiplication
$m^\CF$ \eqref{multiplication}, which reproduces the coordinate
algebra of $\sspdef$, e.g.\ we have
\begin{equation}
\begin{aligned}
\{\theta^{i\alpha},\theta^{j\beta}\}_\star&\ :=\
m^\CF(\theta^{i\alpha}\otimes\theta^{j\beta})+m^\CF(\theta^{j\beta}\otimes\theta^{i\alpha})\\&\
=\ \theta^{i\alpha}\theta^{j\beta}+\frac{\hbar}{2}
C^{i\alpha,j\beta}
+\theta^{j\beta}\theta^{i\alpha}+\frac{\hbar}{2}
C^{j\beta,i\alpha}\\&\ =\ \hbar C^{i\alpha,j\beta}~.
\end{aligned}
\end{equation}
Thus, we have constructed a representation of the Euclidean super
Poincar{\'e} algebra on $\sspdef$ by employing $\CS_\star$, thereby
making twisted supersymmetry manifest.

\section{Applications}\label{sec:apps}

We saw in the above construction of the twisted Euclidean super
Poincar{\'e} algebra that our description is equivalent to the
standard treatment of Moyal-Weyl-deformed superspace. We can
therefore use it to define field theories via their Lagrangians,
substituting all products by star products, which then will be
invariant under twisted super Poincar{\'e} transformations. This can
be directly carried over to quantum field theories, replacing the
products between operators by star products. Therefore, twisted
super Poincar{\'e} invariance, in particular twisted supersymmetry,
will always be manifest.

As a consistency check, we want to show that the tensor
$C^{i\alpha,j\beta}:=\{\theta^{i\alpha},\theta^{j\beta}\}_{\star}$
is invariant under twisted super Poincar{\'e} transformations before
tackling more advanced issues. Furthermore, we want to relate the
representation content of the deformed theory with that of the
undeformed one by scrutinizing the Casimir operators of this
superalgebra. Eventually, we will turn to supersymmetric
Ward-Takahashi identities and their consequences for
renormalizability.

\subsection{Invariance of $C^{i\alpha,j\beta}$}

The action of the twisted supersymmetry charge on
$C^{i\alpha,j\beta}$ is given by
\begin{equation}
\begin{aligned}
\hbar Q^\CF_{k\gamma} C^{i\alpha, j\beta} &\ =\
Q^\CF_{k\gamma}\left(\{\theta^{i\alpha},\theta^{j\beta}\}_\star\right)\\
&\ :=\
m^\CF\circ\left(\Delta^\CF(Q_{k\gamma})(\theta^{i\alpha}\otimes\theta^{j\beta}+
\theta^{j \beta} \otimes \theta^{i \alpha})\right)\\
&\ =\
m^\CF\circ\left(\Delta(Q_{k\gamma})(\theta^{i\alpha}\otimes\theta^{j\beta}+
\theta^{j \beta} \otimes \theta^{i \alpha})\right)\\ &\ =\ m \circ
\CF^{-1} ( \delta_k^i \delta_{\gamma}^{\alpha} \otimes \theta^{j
\beta} +\delta_k^j \delta_{\gamma}^{\beta} \otimes \theta^{i
\alpha} - \theta^{i \alpha} \otimes \delta_k^j
\delta_{\gamma}^{\beta} - \theta^{j \beta} \otimes \delta_k^i
\delta_{\gamma}^{\alpha})\\ &\ =\ m ( \delta_k^i
\delta_{\gamma}^{\alpha} \otimes \theta^{j \beta} +\delta_k^j
\delta_{\gamma}^{\beta} \otimes \theta^{i \alpha} - \theta^{i
\alpha} \otimes \delta_k^j \delta_{\gamma}^{\beta}
- \theta^{j \beta} \otimes \delta_k^i \delta_{\gamma}^{\alpha})\\
&\ =\ 0~.
\end{aligned}
\end{equation}
Similarly, we have
\begin{equation}
\begin{aligned}
\hbar (\bar{Q}^k_{\gad})^\CF C^{i\alpha,j\beta}&\ =\
m^\CF\circ\left(\Delta^\CF(\bar{Q}^k_{\gad})
(\theta^{i\alpha}\otimes\theta^{j\beta}+ \theta^{j\beta} \otimes
\theta^{i\alpha})\right)\\ &\ =\
m^\CF\circ\left(\Delta(\bar{Q}^k_{\gad})
(\theta^{i\alpha}\otimes\theta^{j\beta}+
\theta^{j\beta} \otimes \theta^{i\alpha})\right)\\
&\ =\ 0~,
\end{aligned}
\end{equation}
and
\begin{equation}
\hbar P^\CF_{\mu \nu} C^{i\alpha,j\beta}\ =\ m^\CF\circ
\left(\Delta(P_{\mu}) (\theta^{i\alpha}\otimes\theta^{j\beta}+
\theta^{j\beta} \otimes \theta^{i\alpha})\right)\ =\ 0~.
\end{equation}
For the action of the twisted rotations and boosts, we get
\begin{equation}
\begin{aligned}
\hbar M^\CF_{\mu\nu} C^{i\alpha,j\beta} &\ =\ m^\CF \circ
\left(\Delta^\CF(M_{\mu \nu})(\theta^{i\alpha}\otimes
\theta^{j\beta}+ \theta^{j\beta} \otimes \theta^{i\alpha})
\right)\\
&\ =\
m\circ\CF^{-1}\CF\Delta(M_{\mu\nu})\CF^{-1}(\theta^{i\alpha}\otimes
\theta^{j\beta}+ \theta^{j\beta} \otimes \theta^{i\alpha})\\
&\ =\ m(\unit\otimes
M_{\mu\nu}+M_{\mu\nu}\otimes\unit)\left((\theta^{i\alpha}\otimes
\theta^{j\beta}+ \theta^{j\beta}\otimes \theta^{i\alpha}) -\hbar
C^{i\alpha,j\beta}\unit\otimes\unit\right)\\
&\ =\ 0~,
\end{aligned}
\end{equation}
where we made use of
$M_{\mu\nu}=\di(y_\mu\dpar_\nu-y_\nu\dpar_\mu)$. Thus,
$C^{i\alpha,j\beta}$ is invariant under the twisted Euclidean
super Poincar{\'e} transformations, which is a crucial check of the
validity of our construction.

\subsection{Representation content}

An important feature of noncommutative field theories was
demonstrated recently \cite{Chaichian:2004za,Chaichian:2004yh}:
they share the same representation content as their commutative
counterparts. Of course, one would expect this to also hold for
non-anticommutative deformations, in particular since the
superfields defined, e.g., in \cite{Seiberg:2003yz} on a deformed
superspace have the same set of components as the undeformed ones.

To decide whether the representation content in our case is the
same as in the commutative theory necessitates checking whether
the twisted action of the Casimir operators $P^2= P_{\mu} \star
P^{\mu}$ and $\widetilde{C}^2= \widetilde{C}_{\mu\nu} \star
\widetilde{C}^{\mu\nu}$ on elements of $\CU(\frg)\otimes\CU(\frg)
$ is altered with respect to the untwisted case. But since we have
already shown in (\ref{eq:P}) and (\ref{eq:C2}) that the actions
of the operators  $P_{\mu}$ and $\widetilde{C}_{\mu \nu}$ remain
unaffected by the twist, it follows immediately that the operators
$P^2$ and $\widetilde{C}^2$ are still Casimir operators in the
twisted case. Together with the fact that the representation space
considered as a module is not changed, this proves that the
representation content is indeed the same.

\subsection{Chiral rings and correlation functions}

The chiral rings of operators in supersymmetric quantum field
theories are cohomology rings of the supercharges $Q_{i \alpha}$
and $\bar{Q}^i_\ald$. Correlation functions which are built out of
elements of a single such chiral ring have peculiar properties.

In \cite{Seiberg:2003yz}, the anti-chiral ring was defined and
discussed for non-anticommutative field theories. The chiral ring,
however, lost its meaning: the supersymmetries generated by
$\bar{Q}^i_\ald$ are broken, cf.\ \eqref{defalgebra}, and
therefore the vacuum is expected to be no longer invariant under
this generator. Thus, the $\bar{Q}$-cohomology is not relevant for
correlation functions of chiral operators.

In our approach to non-anticommutative field theory, {\em twisted}
supersymmetry is manifest and therefore the chiral ring can be
treated similarly to the untwisted case as we want to discuss in
the following.

Let us assume that the Hilbert space $\CH$ of our quantum field
theory carries a representation of the Euclidean super Poincar{\'e}
algebra $\frg$, and that there is a unique, $\frg$ invariant
vacuum state $|0\rangle$. Although the operators $Q_{i\alpha}$ and
$\bar{Q}^i_\ald$ are not related via hermitean conjugation when
considering supersymmetry on Euclidean spacetime, it is still
natural to assume that the vacuum is annihilated by both
supercharges. The reasoning for this is basically the same as the
one employed in \cite{Seiberg:2003yz} to justify the use of
Minkowski superfields on Euclidean spacetime: one can obtain a
complexified supersymmetry algebra on Euclidean space from a
complexified supersymmetry algebra on Minkowski
space.\footnote{One can then perform all superspace calculations
and impose suitable reality conditions on the component fields in
the end.} Furthermore, it has been shown that in the
non-anticommutative situation, just as in the ordinary undeformed
case, the vacuum energy of the Wess-Zumino model is not
renormalized \cite{Britto:2003aj}.

We can now define the ring of chiral and anti-chiral operators by
the relations
\begin{equation}
\lsc \bar{Q},\CO \rsc_\star\ =\ 0 \eand \lsc
Q,\bar{\CO}\rsc_\star\ =\ 0~,
\end{equation}
respectively. In a correlation function built from chiral
operators, $\bar{Q}$-exact terms, i.e.\ terms of the form $\lsc
\bar{Q},A\rsc_\star$, do not contribute as is easily seen from
\begin{equation}\label{WardIdentity1}
\begin{aligned}
\langle \lsc \bar{Q},A
\rsc_\star\star{\CO}_1\star\ldots\star{\CO}_n\rangle\ =\ & \langle
\lsc
\bar{Q},A\star{\CO}_1\star\ldots\star{\CO}_n\rsc_\star\rangle\pm\langle
A\star\lsc \bar{Q},{\CO}_1
\rsc_\star\star\ldots\star{\CO}_n\rangle\\&\pm\ldots\pm\langle
A\star {\CO}_1\star\ldots\star\lsc \bar{Q},{\CO}_n\rsc_\star \rangle\\
\ =\ &\langle \bar{Q}
A\star{\CO}_1\star\ldots\star{\CO}_n\rangle\pm\langle
A\star{\CO}_1\star \ldots\star{\CO}_n \star\bar{Q}\rangle\ =\ 0~,
\end{aligned}
\end{equation}
where we used that $\bar{Q}$ annihilates both $\langle 0|$ and
$|0\rangle$. Therefore, the relevant operators in the chiral ring
consist of the $\bar{Q}$-closed modulo the $\bar{Q}$-exact
operators. The same argument holds for the anti-chiral ring after
replacing $\bar{Q}$ with $Q$, namely
\begin{equation}\label{WardIdentity2}
\begin{aligned}
\langle \lsc Q,A
\rsc_\star\star\bar{\CO}_1\star\ldots\star\bar{\CO}_n\rangle\ =\ &
\langle \lsc
Q,A\star\bar{\CO}_1\star\ldots\star\bar{\CO}_n\rsc_\star\rangle\pm\langle
A\star\lsc Q,\bar{\CO}_1
\rsc_\star\star\ldots\star\bar{\CO}_n\rangle\\&\pm\ldots\pm\langle
A\star
\bar{\CO}_1\star\ldots\star\lsc Q,\bar{\CO}_n\rsc_\star \rangle\\
\ =\ &\langle Q
A\star\bar{\CO}_1\star\ldots\star\bar{\CO}_n\rangle\pm\langle
A\star\bar{\CO}_1\star \ldots\star\bar{\CO}_n\star Q\rangle\ =\
0~.
\end{aligned}
\end{equation}

\subsection{Twisted supersymmetric Ward-Takahashi identities}

The above considered properties of correlation functions are
particularly useful since they imply a twisted supersymmetric
Ward-Takahashi identity: any derivative with respect to the
bosonic coordinates of an anti-chiral operator annihilates a
purely chiral or anti-chiral correlation function. This is due to
the fact that $\dpar\sim \{Q,\bar{Q}\}$ and therefore any
derivative gives rise to a $Q$-exact term, which causes an
anti-chiral correlation function to vanish. Analogously, the
bosonic derivatives of chiral correlation functions vanish. Thus,
the correlation functions are independent of the bosonic
coordinates, and we can move the operators to a far distance of
each other. This causes the correlation function to
factorize\footnote{This observation has first been made in
\cite{Novikov:1983ee}.}:
\begin{equation}\label{clustering}
\langle\bar{\CO}_1(x_1)\star\ldots\star\bar{\CO}_n(x_n)\rangle\ =\
\langle\bar{\CO}_1(x^\infty_1)\rangle\star\ldots\star\langle\bar{\CO}_n(x^\infty_n)\rangle~.
\end{equation}
and such a correlation function therefore does not contain any
contact terms. This phenomenon is called {\em clustering} in the
literature.

Another direct consequence of \eqref{WardIdentity1} is the
holomorphic dependence of the chiral correlation functions on the
coupling constants, i.e.
\begin{equation}
\der{\bl}\langle\CO_1\star\ldots\star\CO_n\rangle\ =\ 0~.
\end{equation}
As an illustrative example for this, consider the case of a
$\CN=1$ superpotential `interaction'
\begin{equation}
\CL_{\mathrm{W}}\ =\ \int\dd^2\theta \lambda \Phi+
\int\dd^2\bar{\theta}\bl\bar{\Phi}~,
\end{equation}
where
$\Phi=\phi(y)+\sqrt{2}\theta^\alpha\psi_\alpha(y)+\theta^2F(y)$ is
a chiral superfield and one of the supersymmetry
transformations is given by
$\{Q_\alpha,\psi_\beta\}=\eps_{\alpha\beta}F$. Then we have
\begin{equation}\label{WardIdentity3}
\begin{aligned}
\der{\bl}\langle\CO_1\star\ldots\star\CO_n\rangle&\ =\ \int\dd^4
y\dd^2\bar{\theta}\langle\CO_1\star\ldots\star\CO_n\star\bar{\Phi}\rangle\
=\ \int\dd^4 y\langle\CO_1\star\ldots\star\CO_n F\rangle\\ &\ =\
\int\dd^4
y\langle\CO_1\star\ldots\star\CO_n\{\bar{Q}_\ald,\bar{\psi}^\ald\}\rangle\
=\ 0~.
\end{aligned}
\end{equation}

\subsection{Comments on non-renormalization theorems}

A standard perturbative non-renormalization theorem\footnote{For
more details and a summary of non-renormalization theorems, see
\cite{buchbinder}.} for $\CN=1$ supersymmetric field theory states
that every term in the effective action can be written as an
integral over $\dd^2\theta\dd^2\bar{\theta}$. It has been shown in
\cite{Britto:2003aj} that this theorem also holds in the
non-anticommutative case. The same is then obviously true in our
case of twisted and therefore unbroken supersymmetry, and the
proof carries through exactly as in the ordinary case.

Furthermore, in a supersymmetric non-linear sigma-model, the
superpotential is not renormalized. A nice argument for this fact
was given in \cite{Seiberg:1993vc}. Instead of utilizing Feynman
diagrams and supergraph techniques, one makes certain naturalness
assumptions about the effective superpotential. These assumptions
turn out to be strong enough to enforce a non-perturbative
non-renormalization theorem.

In the following, let us demonstrate this argument in a simple
case, following closely \cite{Argyres}. Take a non-linear sigma
model with superpotential
\begin{equation}
\CW=\tfrac{1}{2}m\Phi^2+\tfrac{1}{3}\lambda \Phi^3~,
\end{equation}
where $\Phi=\phi+\sqrt{2}\theta\psi+\theta\theta F$ is an ordinary
chiral superfield. The assumptions we impose on the effective
action are the following:
\begin{itemize}
\setlength{\itemsep}{-1mm}
\item[$\triangleright$] Supersymmetry is also a symmetry of the
effective superpotential.
\item[$\triangleright$] The effective superpotential is holomorphic
in the coupling constants.
\item[$\triangleright$] Physics is smooth and regular under the possible
weak-coupling limits.
\item[$\triangleright$] The effective superpotential preserves the
$\sU(1)\times \sU(1)_R$ symmetry of the original superpotential
with charge assignments $\Phi:(1,1)$, $m:(-2,0)$,
$\lambda:(-3,-1)$ and $\dd^2\theta:(0,-2)$.
\end{itemize}
It follows that the effective superpotential must be of the form
\begin{equation}\label{eq:Weff}
\CW_{\mathrm{eff}}=m\Phi
W\left(\frac{\lambda\Phi}{m}\right)=\sum_i a_i \lambda^i
m^{1-i}\Phi^{i+2}~,
\end{equation}
where $W$ is an arbitrary holomorphic function of its argument. Regularity of
physics in the two weak-coupling limits $\lambda\rightarrow 0$ and
$m\rightarrow 0$ then implies that $\CW_{\mathrm{eff}}=\CW$.

To obtain an analogous non-renormalization theorem in the
non-anticommutative setting, we make similar assumptions about the
effective superpotential as above. We start from
\begin{equation}
\CW_{\star}=\tfrac{1}{2}m\Phi\star\Phi+\tfrac{1}{3}\lambda
\Phi\star\Phi\star\Phi~,
\end{equation}
and assume the following:
\begin{itemize}
\setlength{\itemsep}{-1mm}
\item[$\triangleright$] {\em Twisted} supersymmetry is a symmetry of the
effective superpotential. Note that this assumption is new
compared to the discussion in \cite{Britto:2003aj}. Furthermore,
arguments substantiating that the effective action can always be written
in terms of star products have been given in
\cite{Britto:2003aj2}.
\item[$\triangleright$] The effective superpotential is holomorphic
in the coupling constants. (This assumption is equally natural as
in the supersymmetric case, since it essentially relies on the
existence of chiral and anti-chiral rings, which we proved above
for our setting.)
\item[$\triangleright$] Physics is smooth and regular under the possible
weak-coupling limits.
\item[$\triangleright$] The effective superpotential preserves the
$\sU(1)\times \sU(1)_R$ symmetry of the original superpotential
with charge assignments $\Phi:(1,1)$, $m:(-2,0)$,
$\lambda:(-3,-1)$, $\dd^2\theta:(0,-2)$ and, additionally,
$C^{i\alpha,j\beta}:(0,2)$, $|C|\sim C^{i\alpha,j\beta}
C_{i\alpha,j\beta}: (0,4)$.
\end{itemize}
At first glance, it seems that one can now construct more
$\sU(1)\times \sU(1)_R$-symmetric terms in the effective
superpotential due to the new coupling constant $C$; however, this
is not true. Taking the $C\rightarrow 0$ limit, one immediately
realizes that $C$ can never appear in the denominator of any
term. Furthermore, it is not possible to construct a term
containing $C$ in the nominator, which does not violate the
regularity condition in at least one of the other weak-coupling
limits. Altogether, we arrive at an expression similar to (\ref{eq:Weff})
\begin{equation}
\CW_{\mathrm{eff},\star}=\sum_i a_i \lambda^i m^{1-i}\Phi^{\star
i+2}~,
\end{equation}
and find that $\CW_{\mathrm{eff},\star}=\CW_\star$.

To compare this result with the literature, first note that, in a
number of papers, it has been shown that quantum field theories in
four dimensions with $\CN=\frac{1}{2}$ supersymmetry are
renormalizable to all orders in perturbation theory
\cite{Terashima:2003ri}-\cite{Berenstein:2003sr}. This even
remains true for generic $\CN=\frac{1}{2}$ gauge theories with
arbitrary coefficients, which do not arise as a
$\star$-deformation of $\CN =1$ theories. However, the authors of
\cite{Britto:2003aj,Grisaru:2003fd}, considering the
non-anticommutative Wess-Zumino model we discussed above, add
certain terms to the action by hand, which seem to be necessary
for the model to be renormalizable. This would clearly contradict
our result $\CW_{\mathrm{eff},\star}=\CW_\star$. We conjecture,
that this contradiction is merely a seeming one and that it is
resolved by a resummation of all the terms in the perturbative
expansion. A similar situation was encountered in
\cite{Britto:2003aj2}, where it was found that one could not write
certain terms of the effective superpotential using star products,
as long as they were considered separately. This obstruction,
however, vanished after a resummation of the complete perturbative
expansion and the star product was found to be sufficient to write
down the complete effective superpotential.

Clearly, the above result is stricter than the result obtained in
\cite{Britto:2003aj}, where less constraint terms in the effective
superpotential were assumed. However, we should stress that it is
still unclear to what extend the above assumptions on
$\CW_{\mathrm{eff},\star}$ are really natural. This question
certainly deserves further and deeper study, which we prefer to
leave to future work.

\section{Conclusions and outlook}

We constructed a Drinfeld-twisted Hopf superalgebra and used this
setup to study certain aspects of $\CN =1/2$ supersymmetric
quantum field theories and their $\CN$-extended variants. In
particular, we scrutinized the consequences of this twisting,
i.e.\ the introduction of a twisted (Euclidean) super Poincar{\'e}
symmetry, on various important structures of supersymmetric QFTs,
such as the cohomology ring of chiral operators and related
Ward-Takahashi identities. We found that in our framework, a
twisted version of these notions can be retrieved and can thus be
used to simplify calculations.

Furthermore, we discussed a number of `naturalness' assumptions on
the deformed superpotential which can lead to non-perturbative
non-renormalization theorems similar to those in the $\CN =1$
supersymmetric case. Granted these assumptions, these theorems
bring about many potential simplifications in higher loop
calculations within $\CN = 1/2$ supersymmetric QFT. More work is
needed, however, to clarify the situation here.

Possible future studies might include
Drinfeld-twisted superconformal invariance. Studying
twist-deformed superconformal field theories, following the
discussion of twisted conformal invariance in
\cite{Matlock:2005zn}, could potentially yield further interesting results.
Moreover, the ideas presented above may prove valuable for
introducing a non-anticommutative deformation of supergravity. Building upon
the discussion presented in \cite{Aschieri:2005yw}, one could try to construct
a local version of the twisted supersymmetry. The latter
proposal appears interesting to us and certainly deserves
further investigation.

\section*{Acknowledgements}
We would like to thank Olaf Lechtenfeld, Sebastian Uhlmann, Robert
Wimmer and Martin Wolf for helpful comments on the draft of this
paper. Moreover, C.S.\ would like to thank Petr Kulish for helpful
correspondence on the cocycle condition for Drinfeld twists. The
research of M.I.\ is based upon work supported by the National
Science Foundation under Grant Nos. PHY-0071512 and PHY-0455649,
and with grant support from the US Navy, Office of Naval Research,
Grant Nos. N00014-03-1-0639 and N00014-04-1-0336, Quantum Optics
Initiative. The work of C.S.\ was done within the framework of the
DFG priority program (SPP 1096) in string theory.

\section*{Note added}

After finishing this paper, the article \cite{Zupnik:2005ut}
appeared, in which an analogous construction was discussed for
$\CN=(1,1)$ supersymmetry.

\section*{Appendix}

\renewcommand{\thesubsection}{\Alph{subsection}}
\renewcommand{\theequation}{\Alph{subsection}.\arabic{equation}}
\makeatletter\@addtoreset{equation}{subsection}

\subsection{Definitions}

Recall that a {\em Hopf algebra} is an algebra $H$ over a field
$k$ together with a {\em product} $m$, a {\em unit} $\unit$, a
{\em coproduct} $\Delta:H\rightarrow H\otimes H$ satisfying $
(\Delta\otimes\id)\Delta=(\id\otimes \Delta)\Delta$, a {\em
counit} $\eps:H\rightarrow k$ satisfying
$(\eps\otimes\id)\Delta=\id$ and $(\id\otimes\eps)\Delta=\id$ and
an {\em antipode} $S:H\rightarrow H$ satisfying
$m(S\otimes\id)\Delta=\eps\unit$ and $m(\id\otimes S)\Delta=\eps
\unit$. The maps $\Delta$, $\eps$ and $S$ are unital maps, that is
$\Delta(\unit)=\unit\otimes \unit$, $\eps(\unit)=1$ and
$S(\unit)=\unit$.

A {\em supervector space} is a $\RZ_2$-graded vector space, i.e.,
one can decompose a supervector space into the direct sum of an
even and an odd subspace. If an element $v$ of a supervector space
is contained in the even or the odd subspace, we write
$\tilde{v}=0$ or $\tilde{v}=1$, respectively. A {\em superalgebra}
is a supervector space endowed with an associative multiplication
respecting the grading (i.e.\
$\widetilde{ab}\equiv\tilde{a}+\tilde{b}\mod 2$) and a unit
$\unit$ with $\tilde{\unit}=0$. On superalgebras, we define the
{\em graded commutator} by $\lsc
a,b\rsc:=ab-(-1)^{\tilde{a}\tilde{b}}ba$.

We fix the following rule for the interplay between the
multiplication and the tensor product $\otimes$ in a superalgebra:
\begin{equation}
(a_1\otimes a_2)(b_1\otimes b_2)\ =\
(-1)^{\tilde{a}_2\tilde{b}_1}(a_1b_1\otimes a_2b_2)~.
\end{equation}

A superalgebra is called a {\em Hopf superalgebra} if it is
endowed with a graded coproduct\footnote{In Sweedler's notation
with $\Delta(a)=\sum a_{(1)}\otimes a_{(2)}$, this amounts to
$\tilde{a}\equiv\tilde{a}_{(1)}+\tilde{a}_{(2)} \mod 2$.} $\Delta$
and a counit $\eps$, both of which are graded algebra morphisms,
i.e.\
\begin{equation}
\Delta(ab)\ =\ \sum
(-1)^{\tilde{a}_{(2)}\tilde{b}_{(1)}}a_{(1)}b_{(1)}\otimes
a_{(2)}b_{(2)}\eand \eps(ab)\ =\ \eps(a)\eps(b)~,
\end{equation}
and an antipode $S$ which is a graded algebra anti-morphism, i.e.\
\begin{equation}
S(ab)\ =\ (-1)^{\tilde{a}\tilde{b}}S(b)S(a)~.
\end{equation}
As usual, one furthermore demands that $\Delta$, $\eps$ and $S$
are unital maps, that $\Delta$ is coassociative and that $\eps$
and $S$ are counital. For more details, see \cite{Brouder:2004}
and references therein.

\subsection{Extended graded Baker-Campbell-Hausdorff formula}

First, note that $\de^{A\otimes B}\de^{-A\otimes B}$ is indeed
equal to $\unit\otimes\unit$ for any two elements $A,B$ of a
superalgebra. This is clear for $\tilde{A}=0$ or $\tilde{B}=0$.
For $\tilde{A}=\tilde{B}=1$ it is most instructively gleaned from
\begin{equation}
\left(\unit\otimes\unit+A\otimes B-\tfrac{1}{2}A^2\otimes
B^2+\ldots\right)\left(\unit\otimes\unit-A\otimes
B-\tfrac{1}{2}A^2\otimes B^2-\ldots\right)\ =\ \unit\otimes\unit~.
\end{equation}

Now, for elements $A_I,B_J,D$ of a graded algebra, where the
parities of the elements $A_I$ and $B_J$ are all equal
$\tilde{A}=\tilde{A}_I=\tilde{B}_J$ and $\lsc A_I,A_J\rsc=\lsc
B_I,B_J\rsc=0$, we have the relation
\begin{align}\label{bchformula}
&\de^{C^{IJ}A_I\otimes B_J} \left(D\otimes \unit\right)
\de^{-C^{KL}A_K\otimes B_L}\\\nonumber&\hspace{1cm} \ =\
\sum_{n=0}^\infty
\frac{(-1)^{n\tilde{A}\tilde{D}+\frac{n(n-1)}{2}\tilde{A}}}{n!}C^{I_1J_1}\ldots
C^{I_nJ_n}\lsc A_{I_1},\lsc\ldots\lsc A_{I_n},D\rsc\rsc\rsc
\otimes B_{J_1}\ldots B_{J_n}~.
\end{align}
{\em Proof:} To verify this relation, one can simply adapt the
well-known iterative proof via a differential equation. First note
that
\begin{equation}
\de^{\lambda C^{IJ}A_I\otimes B_J}(C^{KL}A_K \otimes B_L)\ =\
(C^{KL}A_K\otimes B_L)\de^{\lambda C^{IJ}A_I\otimes B_J}~.
\end{equation}
Then define the function
\begin{equation}
F(\lambda)\ :=\ \de^{\lambda C^{IJ}A_I\otimes B_J}(D\otimes
1)\de^{-\lambda C^{KL}A_K\otimes B_L}~,
\end{equation}
which has the derivative
\begin{align}
\dder{\lambda}F(\lambda)\ =\ (C^{MN}A_M\otimes B_N)&\de^{\lambda
C^{IJ}A_I\otimes B_J}(D\otimes 1)\de^{-\lambda C^{KL}A_K\otimes
B_L}\\\nonumber&-\de^{\lambda C^{IJ}A_I\otimes B_J}(D\otimes
1)\de^{-\lambda C^{KL}A_K\otimes B_L}(C^{MN}A_M\otimes B_N)~.
\end{align}
Thus, we have the identity $\dder{\lambda}
F(\lambda)=[(C^{MN}A_M\otimes B_N),F(\lambda)]$, which, when
applied recursively together with the Taylor formula, leads to
\begin{equation}
F(1)\ =\ \sum_{n=0}^\infty \frac{1}{n!}
\left[C^{I_1J_1}A_{I_1}\otimes
B_{J_1}\left[\ldots\left[C^{I_nJ_n}A_{I_n}\otimes
B_{J_n},D\otimes\unit\right]\ldots\right]\right]~.
\end{equation}
Also recursively, one easily checks that
\begin{align}
&\left[C^{I_1J_1}A_{I_1}\otimes
B_{J_1}\left[\ldots\left[C^{I_nJ_n}A_{I_n}\otimes
B_{J_n},D\otimes\unit\right]\ldots\right]\right]\\\nonumber&\hspace{2cm}\
=\ (-1)^{\tilde{A}\tilde{D}}(-1)^\kappa C^{I_1J_1}\ldots
C^{I_nJ_n}\lsc A_{I_1},\lsc\ldots\lsc A_{I_n},D\rsc\rsc\rsc
\otimes B_{J_1}\ldots B_{J_n}~,
\end{align}
where $\kappa$ is given by
$\kappa=(n-1)\tilde{A}+(n-2)\tilde{A}+\ldots+\tilde{A}$.
Furthermore, we have
\begin{equation}
(-1)^\kappa\ =\ (-1)^{n^2-\sum_{i=1}^n i}\ =\
(-1)^{n^2+\sum_{i=1}^n i}\ =\ (-1)^{\frac{n(n-1)}{2}}~,
\end{equation}
which, together with the results above, proves formula
\eqref{bchformula}. This extended graded Baker-Campbell-Hausdorff
formula also generalizes straightforwardly to the case when
$D\otimes \unit$ is replaced by $\unit \otimes D$.

\end{document}